\documentclass[english,twocolumn,prl,superscriptaddress,longbibliography]{revtex4-1}
\usepackage[T1]{fontenc}
\usepackage[latin9]{inputenc}
\usepackage{color}
\usepackage{amsmath}
\usepackage{graphicx}

\usepackage[colorlinks=true,linkcolor=blue,urlcolor=blue,citecolor=blue,anchorcolor=blue]{hyperref}
\makeatletter
\usepackage{amsmath}
\usepackage{subfigure}
\usepackage{graphicx, mathtools}
\usepackage[normalem]{ulem}

\makeatother

\usepackage{babel}
\begin{document}
\newcommand{\be}{\begin{equation}}
\newcommand{\ee}{\end{equation}}
\newcommand{\bn}{\begin{eqnarray}}
\newcommand{\en}{\end{eqnarray}}
\newcommand{\ii}{\'{\i}}
\newcommand{\ca}{\c c\~a}
\newcommand{\uc}{\uppercase}
\newcommand{\tb}{\textbf}
\newcommand{\bw}{\begin{widetext}}
\newcommand{\ew}{\end{widetext}}

\title{ Universal dielectric response across a continuous metal-insulator transition }

\author{Prosenjit Haldar}\email{prosenjith@iisc.ac.in}\affiliation{Centre for Condensed Matter Theory, Department of Physics, Indian Institute of Sciences, Bangalore 560012, India}
\affiliation{Institute of Mathematical Sciences, Taramani, Chennai 600113, India and\\
Homi Bhabha National Institute Training School Complex, Anushakti Nagar, Mumbai 400085, India}
\author{M. S. Laad}\email{mslaad@imsc.res.in}
\author{S. R. Hassan}\email{shassan@imsc.res.in}

\affiliation{Institute of Mathematical Sciences, Taramani, Chennai 600113, India and\\
Homi Bhabha National Institute Training School Complex, Anushakti Nagar, Mumbai 400085, India}

\begin{abstract}
A wide range of disordered materials, including disordered correlated systems, show ``Universal Dielectric Response'' (UDR), followed by a superlinear power-law increase in their optical responses over exceptionally broad frequency regimes.  While extensively used in various contexts over the years, the microscopics underpinning UDR remains controversial.  Here, we investigate the 
optical response of the simplest model of correlated fermions, Falicov-Kimball model (FKM), across the continuous metal-insulator transition 
(MIT) and analyze the associated quantum criticality in detail using cluster extension of dynamical mean field theory (CDMFT).  
  Surprisingly, we find that UDR naturally emerges in the quantum critical region associated with the continuous MIT.  We tie the emergence of these novel features to a many-body orthogonality catastrophe  accompanying the onset of strongly correlated {\it electronic glassy dynamics} close to the MIT, providing a microscopic realization of Jonscher's time-honored proposal as well as a rationale for similarities in optical responses between correlated electronic matter and canonical glass formers. 
\end{abstract}


\maketitle

\section{Introduction}

   Optical conductivity has long been used to characterize elementary excitations in condensed matter.
Response of matter to $ac$ electromagnetic fields is usually encoded in the complex 
conductivity~\cite{ashcroft-mermin}, $\sigma_{xx}(\omega)=\sigma'(\omega) + i\sigma''(\omega)$ or the 
complex dielectric constant, $\epsilon(\omega)$, related to each other by 
$\sigma'(\omega)=\omega\epsilon_{0}\epsilon''(\omega)$, where $\epsilon''(\omega)$ quantifies 
the dielectric loss, and $\epsilon_{0}$ is the permittivity of free space.  Optical studies have been 
especially valuable in strongly correlated electronic matter~\cite{imada} and, as a particular example, 
have led to insights into breakdown of traditional concepts in cuprates~\cite{vdM}.

   Such studies have also led to much progress in understanding of complex charge dynamics in disordered 
matter.  In the seventies, pioneering work of Jonscher~\cite{jonscher-nature,jonscher} showed a universal dielectric 
response (UDR) of disordered quantum matter to $ac$ electromagnetic fields, wherein 
$\sigma_{xx}(\omega)\simeq \omega^{\alpha}$ with $\alpha\leq 1$ in the sub-GHz regime.  More recently, Lunkenheimer {\it et al.}~\cite{loidl}
report astonishingly similar responses in a wide class of disordered matter over a more extended energy window: among others, doped, weakly 
and strongly correlated semiconductors exhibit UDR, followed by a superlinear power-law increase in $\sigma(\omega)$, bridging the gap between classical dielectric and infra-red regions.  That this behavior is also 
 common to dipolar and ionic liquids as well as to canonical glass formers. Very recently, materials which belong to the elusive class of spin liquids~\cite{spinliquids} have also been interpreted in terms of UDR: in this case, it is possible that intrinsic disorder, arising from geometric frustration, is implicated in emergence of UDR.  This suggests involvement of a deeper, more fundamental and common element, related to onset of a possibly intrinsic, glassy dynamics in emergence of UDR.  In the context of correlated quantum matter (such as the Mott insulator LaTiO$_{3}$ and  
$Pr_{0.65}(Ca_{0.8}Sr_{0.2})_{0.35}MnO_{0.35}$ (PCSMO)~\cite{loidl}), 
such unconventional ``glassy'' dynamics can emerge near the doping-induced MIT as a consequence of 
substitutional and/or intrinsic disorder due to inhomogeneous electronic phase(s) near the MIT.  On the other hand, early on, Jonscher himself suggested the relevance of many-body processes akin to the seminal Anderson orthogonality catastrophe (AOC) for UDR.  Thus, the link between AOC and an emergent, slow glassy dynamics underlying the electronic processes leading to UDR in disordered, interacting electronic systems remains a challenging and largely unaddressed issue for theory, to our best knowledge.

   Motivated hereby, we investigate these issues by a careful study of the optical response of the FKM. The FKM is the simplest representative model of correlated electrons on a lattice, 
and posesses an exact solution within both DMFT~\cite{freericks} and its cluster extensions~\cite{hrk,laad}.  
Remarkably, it can be solved almost analytically, even in CDMFT~\cite{laad}, leading to enormous computational simplifications in transport studies~\cite{qcmott,qchall,thermo}.  Across a critical $U$, the FKM is known to undergo a $T=0$ continuous MIT of the Hubbard band-splitting type~\cite{freericks}. 

   As found earlier for transport properties, it turns out that precise computation of the optical 
response for the FKM within two-site cellular-DMFT~\cite{ourfirstpaper} is facilitated by the fact that: 
$(i)$ explicit closed-form expressions for the cluster propagators, $G({\bf K},\omega)$ with 
${\bf K}=(0,0),(\pi,\pi)$ greatly reduces computational cost, even in CDMFT, and $(ii)$ the 
cluster-resolved irreducible particle-hole vertex functions are negligibly small and we ignore them in the Bethe-Salpeter equations (BSE)
for all conductivities, thanks to an almost rigorous symmetry argument~\cite{kotliar}: upon a cluster-to-orbital mapping (as is implicit in our mapping of the 
two-site CDMFT to two, ``$S,P$'' channels~\cite{ourfirstpaper}).  In this ``multi-orbital'' scenario, the irreducible vertex corrections entering the 
Bethe-Salpeter equation for the conductivity are still negligibly 
small~\cite{silke}.  Thus, the optical 
conductivity acquires a form similar to DMFT, but with an additional sum over 
cluster momenta (or the two ``S'' and ``P'' channels on the cluster). 

  The rest of this paper is organized as follows:  In Sec.  II we
describe the model we study in this work and the calculation of optical conductivity using Cluster DMFT formalism. In Sec.  III we present our numerical results and analyze $(i)$ Mott-like quantum criticality in optical response using CDMFT and $(ii)$ universal dielectric response across the MIT. We then tie the UDR to an emergent many-body orthogonality
catastrophe in the FKM within our CDMFT approach.  In Sec. IV. we discuss our findings in the context of real materials exhibiting UDR.

\section{General Formulation of Optical Conductivity within Cluster DMFT} 

The Hamiltonian of the spinless FKM model is:
\be
H_{FKM}=-t\sum_{<i,j>}(c_{i}^{\dag}c_{j}+h.c)  -\mu\sum_{i}n_{i,c} + U\sum_{i}n_{i,d}n_{i,c}
\label{Eq1}
\ee
The Hamiltonian describes a band of dispersive fermions ($c,c^{\dag}$) interacting locally via a ``Hubbard'' type 
interaction with dispersionless $d$-fermions.  Since $v_{i}=Un_{i,d}$ is a random (binary) potential in
the symmetry-unbroken phases of the FKM, Eq.~\ref{Eq1} can also be viewed (as is long known~\cite{hubbard-III}) as a model of fermions in a random binary alloy potential.

In recent work~\cite{ourfirstpaper}, we used a cluster extension of DMFT to solve FKM using the Dyson-Schwinger equation of motion technique. Remarkably, the cluster-local Green's function in two-site cluster DMFT is obtained analytically, and reads
\[ \hat{\mathbf{G}}= \left[  \begin{array}{cc}
G_{00}(\omega) & G_{\alpha 0}(\omega) \\
G_{\alpha 0}(\omega) & G_{00}(\omega) \end{array} \right] \] 
where, the matrix element $G_{ij}(\omega)$, with $i,j=0,\alpha$,
\begin{widetext}
\begin{eqnarray}
\label{eq:11}
G_{ij}(\omega) = \left[\frac{1-\langle n_{0d}\rangle-\langle n_{\alpha d}\rangle+\langle n_{0d}n_{\alpha d}\rangle}{\xi_2(\omega)}+\frac{\langle n_{0d}\rangle-\langle n_{0d}n_{\alpha d}\rangle}{\xi_2(\omega)-U} \right] \left[  \delta_{ij} - \frac{F_2(\omega)}{(t-\Delta_{\alpha 0}(\omega))}(1-\delta_{ij}) \right] \nonumber\\
+\left[\frac{\langle n_{\alpha d}\rangle-\langle n_{0d}n_{\alpha d}\rangle}{\xi_1(\omega)}+\frac{\langle n_{0d}n_{\alpha d}\rangle}{\xi_1(\omega)-U} \right] \left[  \delta_{ij} - \frac{F_1(\omega)}{(t-\Delta_{\alpha 0}(\omega))}(1-\delta_{ij}) \right]
\end{eqnarray}
\end{widetext}
with, $\xi_1(\omega) = (\omega - \Delta_{00}(\omega)-F_1(\omega))$, $\xi_2(\omega) = (\omega - \Delta_{00}(\omega)-F_2(\omega))$ and $F_1(\omega) = \frac{(t-\Delta_{\alpha 0})^2}{\omega - U - \Delta_{00}(\omega)}$,  $F_2(\omega) = \frac{(t-\Delta_{\alpha 0})^2}{\omega - \Delta_{00}(\omega)}$. 
Where the bath function $\hat{\mathbf{\Delta}}(\omega)$ is related with the local Green's function through a suitable self-consistency condition. The self energy is given as,
\begin{equation}
\hat{\mathbf{\Sigma}}(\omega) = \hat{\mathbf{\mathcal{G}}}^{-1}_0(\omega) - \hat{\mathbf{G}}^{-1}(\omega)
\end{equation} 
with $\hat{\mathbf{\mathcal{G}}}_0(\omega)$ is the Weiss Green's function, $\hat{\mathcal{G}}_0(\omega)=(\omega + \mu)1 - \hat{\mathbf{\Delta}}(\omega)$  
We use the algorithm described in paper~\cite{ourfirstpaper} to find the local Green's function and self energy. In symmetric basis (cluster momentum basis) we can write $G_{S}=(G_{00} + G_{\alpha 0})$ and $G_{P}=(G_{00} - G_{\alpha 0})$ with 'S' and 'P' are even and odd orbital respectively.%  S=(0,0,...) and P=($\pi,\pi,..$).

The optical conductivity is evaluated using the Kubo-Greenwood formalism. In the near-absence of vertex corrections, only the bare bubble, composed from the CDMFT propagators, contributes.
The explicit form of the optical conductivity within Cluster DMFT then reads 
\bn
\sigma'(\omega)&=& \sigma_0 \sum_{\mathbf{K}\in [S,P]} \int_{-\infty}^{\infty} d\epsilon v^2(\epsilon) \rho^{\mathbf{K}}_{0}(\epsilon) \times \nonumber\\
& &\int_{-\infty}^{\infty} d\tilde{\omega} A_{\mathbf{K}}(\epsilon,\tilde{\omega})A_{\mathbf{K}}(\epsilon,\tilde{\omega}+\omega)\frac{f(\tilde{\omega})-f(\tilde{\omega}+\omega)}{\omega}\nonumber\\
\label{Eq2}
\en
with
\be
A_{\mathbf{K}}(\epsilon,\omega)=Im\Big[\frac{1}{\omega+\mu-\epsilon-\Sigma_{\mathbf{K}}(\omega)} \Big]
\ee
Here, $\rho^{\mathbf{K}}_{0}(\epsilon)$ is non-interacting spectral function of the even and odd orbitals and $f(\omega)$ is the Fermi distribution. This simplification allows comprehensive study of the optical response of the FKM within  CDMFT, which we now describe. 

\section{Results and Discussion}
We consider the Bethe lattice with the halfbandwidth of the conduction electron ($c-$ fermions) as unity ($2t=1$). We define the short-range order paremeter $f_{0\alpha}$ as, $f_{0\alpha}=\langle n_{0d}n_{\alpha d}\rangle-\langle n_{0d}\rangle\langle n_{\alpha d}\rangle$.

\subsection{Quantum Criticality near MIT}
We exhibit the real part of the optical conductivity near and across the MIT 
\begin{figure}
\includegraphics[width=1.\columnwidth , height= 
1.\columnwidth]{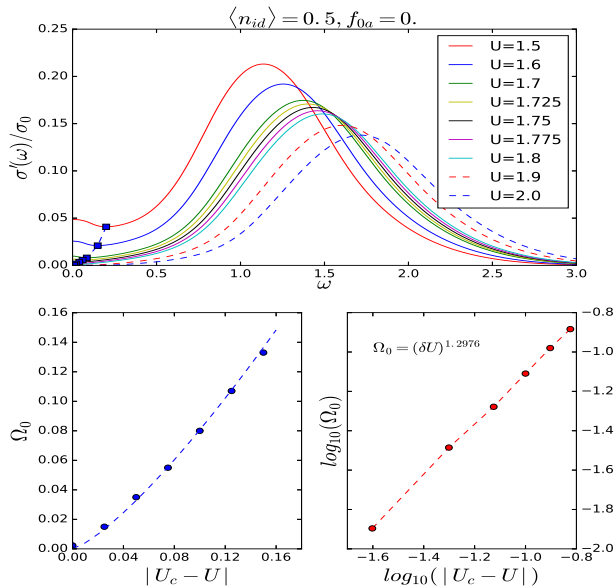} 
\caption{(Color online) Optical Conductivity of the completely random ($f_{0\alpha}=0$) FKM within two-site CDMFT, showing its evolution with $U$ at temperature $T\rightarrow 0$.  The MIT occurs at $U_{c}=1.8$.  Blue symbols show how an emergent 
scale, $\Omega_{0}(U)$, associated with a smooth crossover between metallic and insulating states, collapses at the Mott transition ($U=1.8$) as $(\delta U)^{\nu}$ with $\nu=1.29$, close to $4/3$
(see text).}
\label{fig:fig1}
\end{figure}
($1.6\leq U\leq 2.0$), computed from Eq.~\ref{Eq2} as a function of $U$ for $(a)$ the completely disordered 
case (short-range order parameter $f_{0\alpha}=0$ in our earlier work~\cite{ourfirstpaper}) in the upper panel of Fig.~\ref{fig:fig1} and $(b)$ the  short-range ordered case  ($f_{0\alpha}\neq 0$) in Fig.~\ref{fig:fig2}.  Several features stand out 
\begin{figure}
\includegraphics[width=1.0\columnwidth , height= 
1.0\columnwidth]{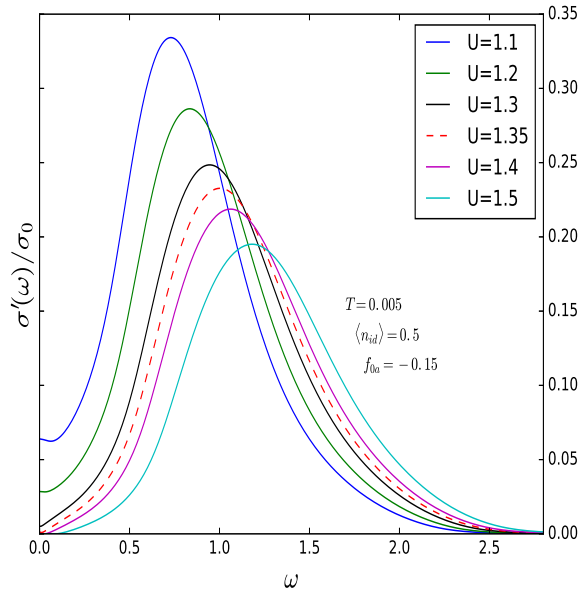} 
\caption{(Color online) The real part of optical conductivity of the FKM with anti-ferro (AF) short-range order ($f_{0\alpha}=-0.15$) within two site CDMFT close to the MIT (1 < U < 1.5). The critical curve at which the MIT occurs corresponds to $(U)_c = 1.35$ (red dotted line).}
\label{fig:fig2}
\end{figure}
clearly: In Case $(a)$, (1) $\sigma'(\omega)$ shows an incoherent low-energy bump centered at 
$\omega=0$, whose weight decreases continuously as the MIT is approached (at $U=1.8$).  It is important 
to note that $(i)$ there is no low-energy Drude component in $\sigma'(\omega)$, since the CDMFT propagators have no pole structure~\cite{ourfirstpaper}, and $(ii)$ as expected, low-energy spectral weight is continuously transferred from the bad-metallic and mid-infra-red (MIR) regions to high energies $O(U)$ across the MIT.  This is characteristic of a correlation-driven MIT
 and the continuous depletion of low-energy weight is a consequence of the continuous MIT in the FKM driven by increasing $U$.  
In Fig~\ref{fig:fig2}, we exhibit the effect of ``anti-ferro alloy'' (AF-A) SRO.  Apart from the fact that the MIT 
now occurs at $(U)\simeq 1.35$~\cite{ourfirstpaper}, the above features seem to be reproduced in this case 
as well.  Looking more closely, however, we see marked changes in the low- and mid-infra-red energy range:
the ``bad metallic'' bump centered at $\omega=0$ is suppressed by SRO, and $\sigma'(\omega)$ rises 
faster with $\omega$ in the MIR, showing up the emergence of a low-energy pseudogap.  These changes are to be expected, since AF-A SRO reduces the effective kinetic energy, increases the effective $U$, leading to reduction of low-energy spectral weight and a low-energy pseudogap in optics.
\begin{figure}
\includegraphics[width=1.\columnwidth , height= 
1.\columnwidth]{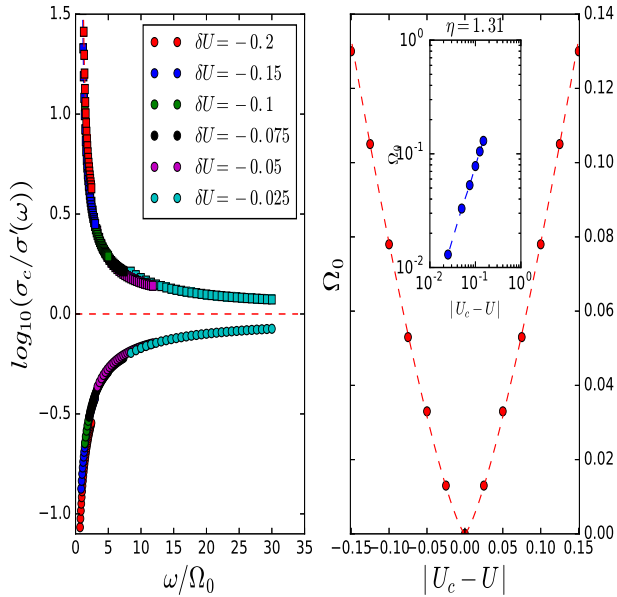} 
\caption{(Color online) Clean quantum critical scaling of the optical conductivity across the Mott QCP, as shown by the fact that log$(\sigma_{c}/\sigma'(\omega))$ versus $\omega/\Omega_{0}(U)$ for the metal and insulator phases falls on two universal "master" curves. $\sigma_{c}$ is the optical conductivity at the critical $U$ i.e. separatrix. We estimate $\Omega_{0}(\delta U)\simeq (\delta U)^{\eta}$ with $\eta=1.31$ in very good accord with $\nu=4/3$ from earlier work~\cite{qcmott}.}
\label{fig:fig3}
\end{figure}
   
  A closer look at Fig.~\ref{fig:fig1} reveals very interesting features.  We uncover a crossover scale ($\Omega_{0}(U)$), separating ``metallic'' and ``insulator-like'' behaviours in $\sigma'(\omega)$.  As expected, it collapses at the MIT: interestingly, we find
\begin{figure}
\includegraphics[width=1.\columnwidth , height= 1.\columnwidth]{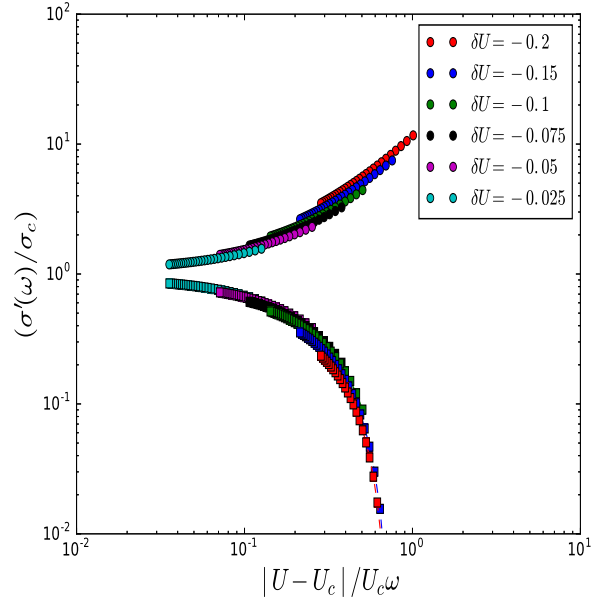} 
\caption{(Color online)  Clean quantum criticality as revealed in the scaling behaviour of the CDMFT optical conductivity, $\sigma'(\omega)/\sigma_{c}$ versus $y=|U-U_{c}|/U_c\omega$.  Since the localization length, $\xi\simeq |U-U_{c}|^{-4/3}$ (see text), this implies that $\sigma'(\omega)/\sigma_{0}=F(\omega\xi^{1/z\nu})$.  This is a manifestation of the "Time Temperature Superposition Principle"~\cite{jonscher} following from Jonscher's UDR.}
\label{fig:fig4}
\end{figure}
$\Omega_0(\delta U)\simeq (\delta U)^{1.29}$, quite close to $\nu=4/3$ found in earlier work~\cite{qcmott}.  This also motivates us to investigate underlying quantum-criticality
in optical response.  In Fig~\ref{fig:fig3}, we show that
log$(\sigma_{c}/\sigma'(\omega))$ plotted versus $\omega/\Omega_{0}(U)$ (the latter taken from Fig.~\ref{fig:fig1}) indeed reveals clean quantum -critical scaling: the insulating ($I$) and metallic ($M$) data fall on two master curves, and the beautiful mirror symmetry relating the two testifies to the unambiguous manifestation of the ``Mott'' QCP in optical response.  Further, we also find
that $\Omega_{0}(\delta U)\simeq c|\delta U|^{\eta}$ with $\eta=1.3\simeq 4/3$, in excellent accord with both Fig.~\ref{fig:fig1} and our previous study.  Using our earlier result $\xi\simeq (U-U_{c})^{-\nu}$ with $\nu=4/3$ and $z=1$, we thus expect that $\sigma'(\omega)/\sigma_{c}$ should scale as $y=|U-U_{c}|/U_c\omega=1/\omega\xi^{1/z\nu}$, {\it i.e}, that $\sigma'(\omega)/\sigma_{c}=F(\omega\xi^{1/z\nu})$.  This is indeed adequately borne out in Fig.~\ref{fig:fig4}, for both $M$ and $I$ phases.  This is a manifestation of the ''Time-Temperature Superposition Principle''~\cite{jonscher}, expressible as a scaling law, $\frac{\sigma'(\omega)}{\sigma_{c}}=F(\frac{\omega}{\Omega_{0}})$
with $F$ a $T$-independent scaling function and $\Omega_{0}(U)$ a scaling parameter corresponding to the onset of conductivity dispersion, precisely as found here.  The variation of $\Omega_{0}$ with $U$ reflects the non-trivial interplay between itinerance (hopping) and Mott-like localization in the FKM.
In analogy with the parameter $T_{0}(\delta U)\simeq c(\delta U)^{z\nu}$ for the $dc$ transport criticality~\cite{qcmott}, $\Omega_{0}$ also scales like 
$(\delta U)^{z\nu}$.  Finally, the fact that $\Omega_{0}(\delta U)\simeq (\delta U)^{z\nu}$ in {\it both} M- and I-phases reflects the fact, alluded to in earlier work~\cite{dobrosavljevic,qcmott}, that the basic electronic processes governing the I-phase are also relevant deep into the M-phase.

\subsection{Unuiversal Dielectric Response}
  Having shown a novel Mott-like quasi-local quantum criticality, we now turn to the UDR near the MIT.
\begin{figure}
\includegraphics[width=1.\columnwidth , height= 
1.\columnwidth]{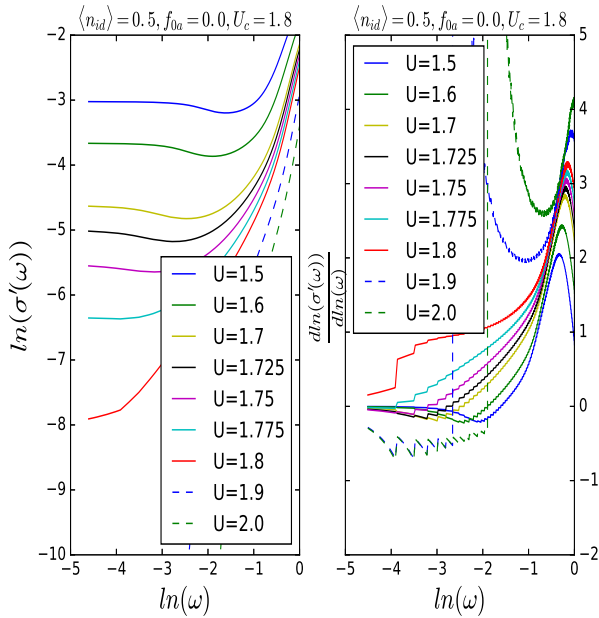} 
\caption{(Color online) Real part of the optical conductivity versus frequency, $\omega$, plotted on a log-log scale to facilitate direct comparison with data for disordered and correlated electronic systems from Lunkenheimer {\it et al.}~\cite{loidl}.  Very good accord is clearly seen.  More importantly, the crossover from the $dc$ limit to UDR around ln$(\omega)\simeq -3$ close to the Mott QCP (for $U\simeq 1.8$) is also revealed, showing that UDR emerges in the quantum critical region associated with the continuous MIT.}
\label{fig:fig5}
\end{figure}  
  Since the FKM is isomorphic to the binary-alloy Anderson disorder problem, we inquire how CDMFT
 performs in the context of the remarkable universality in dielectric response in disordered quantum 
 matter alluded to before~\cite{loidl}.  In Fig.~\ref{fig:fig5}, Fig.~\ref{fig:fig6}, we show log$\sigma'(\omega)$ 
 and the dielectric loss, log$(\sigma'(\omega)/\omega)$ versus log$(\omega)$ as functions of $U$ to facilitate meaningful
 comparison with data of Lunkenheimer {\it et al.}  It is indeed quite remarkable that {\it all} the 
 basic features reported for disordered quantum matter are faithfully reproduced by our CDMFT calculation.  
 Specifically, $(i)$ for $1.5<(U)<1.8$, a ``dc'' conductivity regime at lowest energy (up to $10^{-4}-10^{-3}$) smoothly goes over to a sublinear-in-$\omega$ regime (UDR, in the region $(10^{-2}-10^{-1})$) 
followed by a superlinear-in-$\omega$ regime (around $(10^{-1})$, connecting up smoothly into the 
``boson'' peak.  These regimes are especially visible around $(U)=1.8$, precisely where the MIT occurs.
$(ii)$ Moreover, corroborating behavior is also clearly seen in Fig.~\ref{fig:fig6}, where we exhibit the
dielectric loss function vs $\omega$ on a log-log scale.  It is clearly seen that a shallow minimum 
separates the UDR and superlinear regimes at approximately ln$(\omega) =-0.8$ (in the meV region) in the 
very bad-metallic state close to the MIT.  This is in excellent accord with results for both LaTiO$_{3}$ 
and PCSMO~\cite{loidl}.  Moreover, the energy dependence of the optical conductivity also seems to be in 
good qualitative accord with data when we compare our results with Figures(1),(2) and (3) of 
Lunkenheimer {\it et al.}  In Fig.~\ref{fig:fig6}(b), we also show that short-range spatial 
correlations do not qualitatively modify these conclusions, attesting to their robustness against finite 
short-range order.  Finally, precisely at the MIT (red curves in Fig.~\ref{fig:fig6}, we unearth a very interesting feature: Im$\epsilon(0,\omega) \simeq \omega^{-\eta}$ with $\eta=0.8, 0.75$ for $f_{0\alpha}=0.0, -0.15$.  Hence, apart from a ``dielectric phase angle'', cot$(\pi\eta/2)$, the real part of the dielectric constant also varies like $\omega^{-\eta}$ as the MIT is approached from the metallic side.  Remarkably, this is a concrete manifestation of the dielectric (or polarization) catastrophe that is expected to occur at a 
QCP associated with a continuous MIT.
\begin{figure}
\includegraphics[width=1.\columnwidth , height= 
1.\columnwidth]{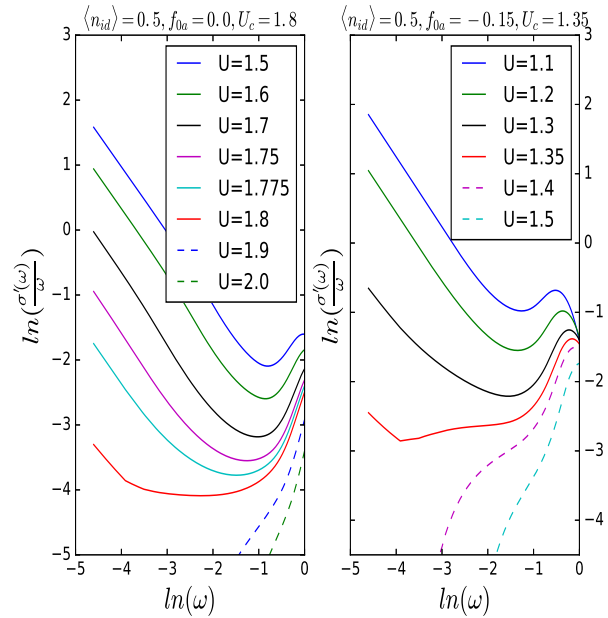} 
\caption{(Color online) The dielectric loss Re$\sigma_{xx}(\omega)/\omega$, plotted versus $\omega$ on a log-log scale to facilitate direct comparison with Lunkenheimer {\it et al.}~\cite{loidl}.  In excellent accord with data on correlated and disordered systems in Ref.$[5]$, a shallow minimum separates the UDR from a superlinear power-law regime around ln$\omega\simeq -1,-2$ as $U$ increases from $1.5$ up to $1.8$, the critical value for the Mott QCP. }
\label{fig:fig6}
\end{figure}

   This wide-ranging accord with data is quite remarkable, and begs a microscopic clarification in terms of basic electronic
processes at work near the MIT.  A phenomenological way is to posit that the universality is linked to 
glassiness~\cite{loidl} as follows: $(i)$ first, our finding~\cite{qcmott} of $\nu=4/3$ (and $z=1$) is characteristic of percolative transport that is naturally  expected to arise in glassy systems.  $(ii)$ it has also been shown~\cite{pastor} (for the disorder problem) that electronic
 glassy behavior {\it precedes} an insulating phase.  Thanks to the mapping between the FKM and a binary-alloy "disorder" problem, we also expect an {\it intrinsic} electronic glassy phase near the continuous MIT in the FKM.  This suggests a close link between UDR and onset of electronic glassy dynamics near the continuous MIT in the FKM.   

  On a more microscopic level, it is {\it very} interesting that Jonscher himself~\cite{jonscher} hinted at relevance of many-body processes at the heart of UDR.  In particular, an explicit parallel with the seminal X-ray edge like physics 
was already (phenomenologically) invoked to account for such features.  It is indeed remarkable that 
such X-ray-edge physics naturally falls out in the DMFT {\it and} CDMFT solution of the FKM in high-$D$~\cite{si-1992}: in DMFT, the mobile ($c$-fermion) 
propagator exhibits a pseudogapped metal going over to an insulator across a 
critical $U=U_{c}$, while the $d$-fermion propagator exhibits a ``X-ray-edge'' 
singular behavior linked to the seminal Anderson orthogonality catastrophe 
(AOC).  In our CDMFT~\cite{ourfirstpaper}, we find similar behaviour: 
$(i)$ a clear correlation-induced pseudogapped metal without Landau 
quasiparticles going over to an insulator at $U_{c}=1.8$, with an anomalous 
self-energy, Im$\Sigma_{loc}^{(c)}(\omega)\simeq |\omega|^{1/3}$ 
and density-of-states $\rho_{c}(\omega)\simeq |\omega|^{1/3}$ at the Mott QCP, 
and $(ii)$ unusual 
power-law behavior of the dynamical charge susceptibility near $U_{c}$, as well as, most importantly,  anomalous energy dependence of the $d$-fermion propagator at long times.  In CDMFT, these features arise precisely from 
the fact that 
$(a)$ the dispersive $c$-fermions interact with the 
$d_{\pm}=(d_{1}\pm d_{2})/\sqrt{2}$ dispersionless fermion modes at the 
{\it intra-cluster} level via $U$, while $(b)$ the $c$-fermions do {\it not} 
hybridize at all with the $d_{\pm}$ localized mode at single-fermion level. 
Physically, the origin of the AOC is that the dynamical 
screening arising from strong intra-cluster interactions in the FKM is 
non-trivially affected by hopping motion of carriers: since the $c_{\pm}$ 
fermions do {\it not} hybridize with the $d_{\pm}$ fermion modes at the 
one-fermion level (there is no term of the form $V(c_{\pm}^{\dag}d_{\pm}+h.c)$ 
on the cluster), this screening is non-trivial, and arises from the ``slow'' 
reaction of the $d_{\pm}$ modes to the ``sudden'' jumps of the $c_{\pm}$-
fermions (the latter occurs on a much shorter time scale of $\hbar/t$ in the 
FKM).   Due to the {\it local} symmetry implied by $[n_{i,d},H_{FKM}]=0$ at 
each $i$, a hopping carrier experiences a sudden change in the local potential 
on the cluster (from $0$ to $U$ and vice versa while hopping), now in the sense
 of a ``sudden local quench''.  Rigorous absence of $c-d$-fermion one-particle 
mixing in the FKM implies lack of heavy-particle recoil in the ``two-impurity''
 cluster problem, leading to ``Kondo destruction'' and generation of an AOC in 
our two-site CDMFT as above, {\it in precise accord with Jonscher's original 
suggestion}. We present the origin of the many-body AOC using an analytical approach in the following subsection.   

\subsection{Cluster Orthogonality Catastrophe in Two-site CDMFT for the FKM}

Here, we present an analytic argument that exposes the venerated orthogonality
catastrophe in our CDMFT approach.  It turns out that this is most conveniently done by taking recourse to the underlying two-impurity problem of our CDMFT, which we now describe.

  The ``two-impurity'' FKM is: 

\begin{eqnarray}
H_{FKM}&=&t(c^{\dag}_0 c_{\alpha}+h.c.) + \sum_{k}\epsilon_{k}c_{k}^{\dag}c_{k} + \nonumber\\
& &t\sum_{k,i=0,\alpha}(e^{ik.R_{i}}c_{i}^{\dag}c_{k}+h.c)+ U\sum_{i=0,\alpha}n_{i,c}n_{i,d}\nonumber\\
\end{eqnarray}

  Introduce the bonding-antibonding fermions, $c_{\pm}=(c_{0}\pm c_{\alpha})/\sqrt{2}, d_{\pm}=(d_{0}\pm d_{\alpha})/\sqrt{2}$.  Then, $H_{FKM}=H_{12}+H_{coupl}+H_{band}$
with

\be
H_{12}=t(n_{c,+}-n_{c,-}) + \frac{U}{2}\sum_{a,a'=\pm}(n_{c,a}n_{d,a'} + c_{a}^{\dag}c_{a'}d_{a}^{\dag}d_{a'}),
\ee

\be
H_{coupl}=\sqrt{2}t\sum_{k}[cos(ka/2)c_{+}^{\dag}c_{k} + isin(ka/2)c_{-}^{\dag}c_{k} + h.c],
\ee

and, 
\be 
H_{band}=\sum_{k}\epsilon_{k}c_{k}^{\dag}c_{k}
\ee

   This ``cluster-to-orbital''  mapping exposes the novel structure of the cluster-local correlations at the Mott QCP.  Specifically, we observe that while the $d_{\pm}$-fermions interact with the $c_{\pm}$-fermions via $U$, they do not hybridize with each other at the one-fermion level ($i.e$, there is no term of the form $V_{\pm}(c_{\pm}^{\dag}d_{\pm} + h.c)$ in the FKM).  Thus, the two-impurity FKM maps rigorously onto a cluster version of the classic problem of recoilless, ``infinite-mass'' $d_{\pm}$ scatterers in a ``Fermi sea'' of the $c_{\pm}$; $i.e$, to the cluster version of the venerated X-ray edge problem.
  One now directly reads off that the spectral function of the $d_{\pm}$ fermions is infra-red singular with an interaction-dependent power-law behavior~\cite{nozieres}:

\be
\rho_{d_{\pm}}(\omega) \simeq \frac{\theta(\omega)}{|\omega|^{1-\eta_{\pm}}}
\ee
where $\eta_{\pm}=(\delta_{\pm}/\pi)^{2}$ and $\pi\delta_{\pm}=$tan$^{-1}(U\rho_{c_{\pm}}(0)\pi)$ is the scattering phase shift.  There will be an additional 
contribution to the scattering phase shift arising from the term 
$\frac{U}{2}\sum_{a,a'}c_{a}^{\dag}c_{a'}d_{a}^{\dag}d_{a'}$, but this will not
 qualitatively change the power-law behavior above.  This many-body orthogonality
 catastrophe will carry over into the self-consistently embedded two-site CDMFT
 solution of the FKM.   

   Interestingly, we thus find that the orthogonality catastrophe and the accompanying breakdown of adiabatic continuity also holds for the case of spatially 
separated recoilless scatterers on the length scale of the cluster.  
Using a different approach, this aspect has also been 
studied earlier~\cite{altshuler}.  Thus, {\it incorporation of inter-site 
correlations between spatially separated scattering centers does not qualitatively modify the 
orthogonality catastrophe}, an interesting result in itself. In modern 
parlance, this means that the vanishing fidelity as well as the anomalous 
long-time behavior of the Loschmidt echo, both manifestations of the 
orthogonality catastrophe~\cite{silva} also hold for spatially separated, sudden
local quenches, a result that may have more widespread applications. 

   Thus, the classic orthogonality catastrophe, arising from the ``sudden local'' but spatially correlated quenches in our two-impurity 
model, is a genuine feature in our CDMFT used in the main text.  This also provides a concrete microscopic rationalization linking the Jonscher UDR to this novel many-body effect, as conjectured early on by Jonscher himself.  

\section{Discussion and Conclusion}

%  Moreover, 
Our findings can be profitably utilized to interpret wider range 
of data on dielectric responses of a wide range of disordered electronic 
matter, e.g, disordered semiconductors, doped Mott insulators, PN junction 
devices~\cite{jonscher}, etc.  In reality, the optical response at low energy 
will now be Re$\sigma_{xx}(\omega)=\sigma_{dc}+\sigma_{0}\omega^{\alpha}$, with
 $0<\alpha <1$.  This directly implies that Im$\sigma_{xx}(\omega)=$tan$(\pi\alpha/2)\sigma_{0}\omega^{\alpha}+\omega\epsilon_{0}\epsilon_{\infty}$, with 
$\sigma_{dc}$ the dc conductivity and $\epsilon_{\infty}$ the bare dielectric 
constant.  The corresponding (dynamic) capacitance and impedance read
$C(\omega)=C_{\infty}+(\sigma_{0}/2\pi)$tan$(\pi\alpha/2)\omega^{\alpha-1}$ 
and $Z^{*}(\omega)\simeq (i\omega\epsilon^{*}(\omega))^{-1}$.  Along with 
causality (Kramers-Kr\"onig relations), UDR implies that the real and imaginary
 parts of the dielectric function (thus of the susceptibility) are related to 
each other by a ``dielectric phase angle'',
$\chi'(\omega)/\chi''(\omega)=$cot$(\pi\alpha/2)$~\cite{jonscher}, 
{\it independent} of $\omega$, in stark contrast to Debye-like relaxation, 
where this ratio equals $\omega\tau$.
Such forms have ben widely used to analyze data in detail for a wide range 
of disordered matter~\cite{jonscher-nature,jonscher,lunkenheimer-batisite} for a long time.  Within 
CDMFT, our findings provide a microscopic rationale for use of these relations.

 Theoretically, it is very interesting that such features appear near a 
correlation-driven MIT in the FKM,
since this is a band-splitting type Mott (rather than pure Anderson 
localization in a disorder model, or a first-order Mott transition in the pure 
{\it Hubbard} model) transition.  It supports
views~\cite{sudip1997,qcmott} that the ``disorder'' problem at strong coupling, 
where $k_{F}l\simeq O(1)$, is characterized by a different Mott-like quantum 
criticality, a view nicely supported by our earlier finding~\cite{qcmott} of 
$\beta(g)\simeq$ ln$(g)$ instead of $\beta(g)\simeq (d-2)-1/g$ even deep in 
the (bad) metallic phase.  This is not
unreasonable, as it is long known~\cite{velicky} that the coherent potential 
approximation (CPA), the best 
mean-field theory for the Anderson disorder problem, is equivalent to the 
Hubbard III band-splitting view 
of the Mott transition (the latter becomes exact for the FKM in 
$d=\infty$~\cite{vollhardt}).  

   As concrete examples on the materials front, we note that $(1)$ various 
aspects of manganite physics have also been successfully modeled by an effective FKM, where the 
$c$-fermions represent effectively spinless fermions (due to strong Hund coupling) strongly scattered by a disordered ``liquid'' of
effectively localized Jahn-Teller polarons~\cite{tvr}.  In this context, it is also interesting to notice that a field-induced percolative MIT is also long known to occur in manganites~\cite{Fath1540}.  Thus, our model can serve as a simplest effective 
model for PCSMO~\cite{loidl}.  Application to LaTiO$_{3}$ would require using a full Hubbard model very 
close to half-filling, where {\it intrinsic} disorder due to inhomogeneous 
phases near the MIT would generally be expected to be relevant.  $(2)$ it is 
also very interesting that related features, namely $(i)$ non-Landau 
quasiparticle (Drude) ``strange'', but {\it infra-red singular} power-law 
optical response and $(ii)$ 
anomalous optical phase angle characterize strange metallicity in 
near-optimally doped cuprates~\cite{vdM}.  One scenario, based on the hidden 
Fermi-liquid idea, posits that an {\it inverse} orthogonality catastrophe also 
underlies~\cite{pwa-nature} such non-Landau Fermi liquid observations in 
cuprates.  However, the microscopics in this case involves 
momentum-selective Mott physics within CDMFT~\cite{haule-kotliar} and such novel responses could be linked to co-existing nodal (``itinerant'') and Mott-localized anti-nodal states.  While more work is certainly needed to cleanly exhibit such features in the cuprate context, it is out of scope of our present study.

   Thus, the central message of our work is that non-perturbative {\it dynamical} effects of strong intrinsic  
scattering in the FKM lead to onset of a many-body AOC.  We find that 
it is this specific aspect that is at the heart of the ``universal'' dielectric response.

%\bibliography{Bibliography}
 % Appendix Title

\end{document}